\documentclass[11pt]{article}
\usepackage{amssymb,citesort}
\begin{document}

\title{The stability limit of the fluid phase of polydisperse sticky spheres}

\author{{\bf Richard P. Sear\footnote{{\large
Permanent address: Department of Physics,
University of Surrey, Guildford, Surrey GU2 5XH, United Kingdom}}}\\
~\\
Department of Chemistry and Biochemistry\\
University of California, Los Angeles\\
Los Angeles, California 90024, U.S.A.\\
email: sear@chem.ucla.edu}

\date{\today}

\maketitle

\begin{abstract}
It has been shown by Stell [{\it J. Stat. Phys.}, {\bf 63}, 1203 (1991)]
that at low temperature monodisperse sticky spheres collapse to
form coexisting close-packed solid and infinitely dilute gases.
We show that polydisperse sticky spheres also collapse and
calculate the collapse temperature.
The polydisperse spheres separate
into fractions with narrower polydispersities which can then solidify.
This is perhaps the first example of a single-peaked polydisperse
mixture phase solidifying and separating.
It implies that a mixture of polydisperse large hard spheres with much smaller
hard spheres does not show fluid--fluid coexistence.
\end{abstract}

\newpage
\section{Introduction}

Sticky spheres are hard spheres with a zero-ranged, `sticky', attraction
\cite{baxter68}. Their phase behaviour is straightforward, if a little
peculiar.
Above a certain temperature $T_{coll}$ they behave as hard spheres
(below close packing \cite{bolhuis94}) and below
this temperature, they phase separate into an infinitely dilute gas
coexisting with a close-packed solid \cite{stell91,bolhuis94}.
As $T_{coll}$ can be determined analytically (see below) the
sticky sphere model is very attractive for
doing simple analytic theory.
So, if we wish to go beyond pure fluids and consider mixtures, and we
wish to use analytic theory then the sticky sphere model is an obvious
choice for generalisation to describe a mixture.
As we will see below, even for a mixture we can determine the
limit of its stability analytically.
Of course, the model is an extreme one but
we think that even so our analytic demonstration of its phase
behaviour is useful.
Our mixture is polydisperse which means that particles of a whole range
of sizes are present. Within this range, particles of all sizes are
found, i.e., the distribution of sizes of the particles is a continuous
function \cite{salacuse82,gualtieri82,briano84}.

In terms of describing experiment,
sticky spheres have been extensively
used as a model for colloidal particles
and proteins, for examples see Refs.
\cite{stell91,borstnik97,rosenbaum96}
and references therein.
Also, the fractionation method of Bibette \cite{bibette91,bibette92}
for emulsions relies on behaviour which is similar to that which we
find.

We find that at low temperature
polydisperse sticky spheres, like monodisperse sticky spheres,
phase separate into an infinitely dilute gas and close packed solid. But
in order to solidify the polydisperse mixture fractionates: the
original polydisperse mixture separates into a number of polydisperse
mixtures with narrower polydispersities, each of which then solidifies.
Thus there is not one solid phase but many solid phases, each
formed from a different range of sizes of spheres.
This fractionation occurs as highly polydisperse spheres cannot
solidify into a single solid phase because a wide range of
sizes of sphere cannot be accommodated within a solid lattice
\cite{pusey87,barrat86,macrae88,bolhuis96}.
As in monodisperse sticky spheres the driving force for solidification
is the lowering of energy when a particle is in contact with,
and so interacting with, the 12 particles
that surround it in a face-centered-cubic or hexagonal-close-packed lattice.

In the next section we examine monodisperse sticky spheres, then in section
3 we generalise the theory to polydisperse spheres.
Finally we discuss our results and point out their relevance to mixtures
of hard spheres.

\section{Monodisperse sticky spheres}

First, we define the sticky sphere potential of Baxter \cite{baxter68}.
It is the spherically symmetric pair potential $u(r)$ defined by
\begin{equation}
u(r)=
\left\{
\begin{array}{ll}
\infty & ~~~~~~ r \le 1\\
-\epsilon & ~~~~~~ 1<r\le 1+\delta \\
0 & ~~~~~~ r > 1+\delta \\
\end{array}\right. ,
\label{monoss}
\end{equation}
in the limit $\delta\rightarrow 0$. We
have defined the diameter of the hard-sphere part of the potential
to be equal to 1.
The effect of the attractive part of the interaction, that between
1 and $1+\delta$, can be assessed using the second virial coefficient
$B_2$, which is
\begin{equation}
B_2=\frac{2\pi}{3}-2\pi\delta\exp(\epsilon/T),
\label{b2}
\end{equation}
where the first term is simply $B_2$ for hard spheres
of unit diameter and the second term
comes from the attractive interactions.
We use temperature units such that Boltzmann's constant $k=1$.
We can easily see that as we cool sticky spheres the attractive interactions
contribute a significant amount to $B_2$ at temperatures
of $\ln(1/\delta)$ and below. In the limit $\delta\rightarrow0$
\begin{equation}
B_2=\left\{
\begin{array}{ll}
\frac{2\pi}{3} & T>T_{B20}\\
-\infty & T<T_{B20}\\
\end{array}\right. ,
\end{equation}
where $T_{B20}$ is
\begin{equation}
\frac{T_{B20}}{\epsilon}=\lim\delta\rightarrow 0
\left(\frac{1}{\ln(1/\delta)}\right).
\label{tb20}
\end{equation}
The temperature $T_{B20}$ has no relevance
for monodisperse sticky spheres \cite{stell91},
however, as we shall see, it may have for polydisperse sticky spheres.
As $B_2$ diverges below $T_{B20}$ Baxter \cite{baxter68,barboy74}
studied sticky spheres in an infinitesimal temperature range at
$T_{B20}$.
However, Stell and Williams \cite{stell91,borstnik97,borstnik94,hemmer90}
found that the fluid phase of monodisperse sticky spheres 
is unstable  {\it at} $T_{B20}$. It is only stable
above a higher temperature $T_{coll}$ \cite{stell91,hemmer90},
when the second virial coefficient is
indistinguishable from that of hard spheres.
We will now determine $T_{coll}$ in a way that we will later generalise to
polydisperse sticky spheres.
The fluid becomes unstable with respect
to a solid phase near close packing. This is demonstrated by
comparing the free energy of a low density fluid phase with that of
the solid. At low density the fluid has a free energy per particle
$a_f$ which is given by that of an ideal gas
\begin{equation}
\frac{a_f}{T}=\ln\rho-1~~~~~~~~~ \rho\ll1,
\label{af}
\end{equation}
where $\rho=N/V$ is the density of particles; $N$ is the number of
particles and $V$ is the volume.
We have neglected a term $\ln\Lambda^3$ where $\Lambda$ is the
de Broglie wavelength of a sticky sphere. This term is irrelevant as
far as the phase behaviour is concerned.
The free energy of the solid phase may be estimated
using a cell theory \cite{buehler51}. The free energy
per particle is, in a cell theory, obtained from the 1-particle
partition function of a particle trapped in a cell
formed from its neighbouring
particles fixed at the positions they occupy in a ideal lattice.
This partition function $q_1$ is
\begin{equation}
q_1=v_f\exp(z\epsilon/(2T))
\label{q1}
\end{equation}
for sticky spheres in a $z$-coordinate lattice with a lattice
constant sufficiently small that the particle interacts with
all $z$ neighbours at any position in the cell.
For sticky spheres this means that the lattice
constant should be less than $1+\delta/2$. This constraint
on the maximum value of the lattice constant implies that
the density must be at least $\rho_{cp}/(1+\delta/2)^3$, where
$\rho_{cp}(z)$ is the maximum possible density of the
$z$-coordinate lattice
With a lattice
constant $\simeq 1+\delta/2$ a sphere can move a distance
$\simeq\delta/2$ in any direction which yields a volume
available to the centre of mass of the particle $\simeq \delta^3/8$.
So, an accurate approximation to $q_1$ is given by
\begin{equation}
q_1= \frac{\delta^3}{8}\exp(z\epsilon/(2T)) ~~~~~~~~
\rho>\frac{\rho_{cp}(z)}{(1+\delta/2)^3},
\end{equation}
which gives a free energy per particle in the solid phase $a_s$ of
\begin{equation}
\frac{a_s}{T}=-\ln q_1=-3\ln\delta - \frac{z\epsilon}{2T} ~~~~~~~~
\rho>\frac{\rho_{cp}(z)}{(1+\delta/2)^3},
\label{as}
\end{equation}
where we have neglected the factor of $\ln 8$ as being negligible.
The free energy of the solid phase is lowest for the
lattice with the highest coordination $z$. Thus a close packed lattice,
either face-centered-cubic or hexagonal-close-packed,
with $z=12$ is
the stable solid phase. From now on we always take $z=12$.
Here we are considering the $\delta\rightarrow 0$ limit and so
Eq. (\ref{as}) is only valid for a solid at close-packing, $\rho=\rho_{cp}$.

In order to determine the stability of the fluid with respect to the
close-packed solid we determine the free energy difference $a_s-a_f$
in the $\delta\rightarrow0$ limit. From Eqs. (\ref{af}) and (\ref{as})
it is, in the limit $\delta\rightarrow 0$,
\begin{equation}
\frac{a_s-a_f}{T}=\left\{
\begin{array}{ll}
\infty & T>T_{coll}\\
-\infty & T<T_{coll}\\
\end{array}\right. ,
\label{aslim}
\end{equation}
where $T_{coll}$ is
\begin{equation}
\frac{T_{coll}}{\epsilon}=\lim\delta\rightarrow 0
\left(\frac{2}{\ln(1/\delta)}\right),
\label{tcoll}
\end{equation}
which is higher than $T_{B20}$, Eq. (\ref{tb20}).
So, below $T_{coll}$ the free
energy is $-\infty$ at close packing in the sticky limit:
the dilute fluid, with free energy given by Eq. (\ref{af}),
then has a higher free energy than the
close-packed solid phase for all
nonzero densities and so is unstable with respect to this phase.
Positivity of the pressure requires that the free energy be an
increasing function of density at constant temperature so
a phase is necessarily unstable with respect to a denser phase
of lower free energy.

At this point we have not quite demonstrated that the fluid phase
is stable above $T_{coll}$ but collapses to
the close-packed solid at this temperature;
we have not considered the effect of the sticky attractions on the
dense fluid phase and the solid phase at densities below close packing.
We will now consider their effect on the solid phase, there should be no
qualitative difference between it and the dense fluid phase.
At densities below $\rho_{cp}/(1+\delta/2)^3$ a sticky sphere cannot
be within $\delta$ of all $z$ of its neighbours
and so at these densities the energy is higher.
If the sphere rattles freely in the cell and the cell is much larger than
$\delta$ across then the energy is close to zero: the solid is
almost indistinguishable from a solid of hard spheres. It is possible
that while keeping the overall density fixed the sticky spheres could form
chains or sheets of spheres in contact.
Forming a chain incurs an entropy cost of $\ln(1/\delta)$ and
releases an energy $\epsilon$.
Thus the free energy change of chain formation becomes negative at
$T_{B20}$, as we would expect given the nature of the second virial
coefficient approximation for the free energy of a fluid \cite{hansen86}.
So, as $T_{coll}>T_{B20}$ chains are not favoured at $T_{coll}$. 
Forming sheets costs
$3\ln(1/\delta)$ and releases an energy $z_s\epsilon/2$ where
$z_s$ is the coordination number of the sheet.
The highest coordination number for a sheet is $z_s=6$, so sheets
are never stable: their entropy is no higher than that of a close-packed
solid but their energy is only half that of the close-packed solid.
So, we conclude that at and above $T_{coll}$ the free energy of sticky spheres
is the same as that of hard spheres at all densities below close
packing and that therefore at $T_{coll}$ sticky spheres at any density
below close packing phase separate into an infinitely dilute
gas coexisting with a close-packed solid \cite{stell91,hemmer90}.
The gas is infinitely dilute as its free energy per particle must be less
than that of the close-packed solid which is $-\infty$.
Above $T_{coll}$ the free energy of the solid diverges to $\infty$
as we approach close packing: the solid behaves as a solid of hard spheres
and the fluid--solid transition is at the same densities as
found for hard spheres.

At higher temperatures,
Bolhuis {\it et al.} \cite{bolhuis94} have shown that {\it at} close packing
there is an expanded-solid--condensed-solid transition which
is analogous to a vapour--liquid transition. This transition persists
up to a critical temperature $T_c/\epsilon={\cal O}(1)$.
However, because this occurs at close packing this transition
is isolated from the fluid phase; the solid which coexists with
the fluid is at much lower density where the attractive interactions
have no effect at a temperature $T/\epsilon={\cal O}(1)$.

\section{Polydisperse sticky spheres}

In a polydisperse mixture of sticky spheres, spheres with a
range of diameters are present \cite{salacuse82,gualtieri82,briano84}.
In the thermodynamic limit there is
a continuous distribution of spheres of
sizes with a density $\rho x(s){\rm d}s$ of spheres of size $s$.
The fraction of spheres of size $s$ is $x(s){\rm d}s$.
The width of the polydispersity is characterised by a width parameter $w$.
The larger is $w$ the broader the distribution of sizes present in
the mixture. In the limit $w\rightarrow0$ we recover a monodisperse
system. Generally, this limit is straightforward but here of course,
we are taking the limit $\delta\rightarrow0$ limit so we must take care
in the limit of small $w$ as then the ratio $w/\delta$ may not
be small and so the phase behaviour will depend on it \cite{stell91}.
In order to recover a monodisperse system the ratio of $w$ to all
other length scales must tend to 0.

Stell \cite{stell91} realised that the phase behaviour
of polydisperse sticky hard spheres depends strongly on the
ratio $r=w/\delta$. For $r\ll1$ then the spheres are effectively
monodisperse and they behave as described in the previous section.
However, in the opposite limit, sufficiently large polydispersity suppresses
the collapse.
However, Stell
did not determine how large must the ratio $r$ be in order to do so.
This is what we do here. We will show that the width
of polydispersity required to stabilise the fluid phase is
a function of temperature.
As the temperature
decreases below the $T_{coll}$ of Eq. (\ref{tcoll}) the width
of the polydispersity required to suppress the collapse to a close-packed
solid and so stabilise the fluid phase increases exponentially.

We will take the range of the attraction to be $\delta$ and the well
depth to be $\epsilon$
for all spheres. This would be appropriate if the attraction
is a depletion attraction induced by the presence of small
spheres \cite{asakura54,vrij76,biben90,raton98} and the polydispersity width
$w$ is much less than the diameter of the spheres.
The functional form of the polydispersity should not matter too much;
we select a very simple form, the hat function.
The function
$x(s)$ is defined to be
\begin{equation}
x(s)=\left\{
\begin{array}{ll}
0 & ~~~~~~ s< 1-w/2 \\
w^{-1} & ~~~~~~ 1-w/2\le s\le 1+w/2 \\
0 & ~~~~~~ s > 1+w/2 \\
\end{array}\right. ,
\label{xs}
\end{equation}
The interaction
$u(r,s,s')$ between a pair of spheres
of species $s$ and $s'$ is
\begin{equation}
u(r,s,s')=\left\{
\begin{array}{ll}
\infty & ~~~~~~ r \le (1/2)(s+s') \\
-\epsilon & ~~~~~~ (1/2)(s+s')<r\le(1/2)(s+s')+\delta \\
0 & ~~~~~~ r > (1/2)(s+s')+\delta \\
\end{array}\right. .
\label{us}
\end{equation}

First, let us consider the low density fluid phase.
The free energy per particle of a polydisperse ideal gas is
\cite{salacuse82,gualtieri82,briano84}
\begin{equation}
\frac{a_f}{T}=\ln\rho+\int x(s)\ln x(s){\rm d}s.
\label{sid}
\end{equation}
For the distribution of Eq. (\ref{xs})
\begin{equation}
\frac{a_f}{T}=\ln\rho -\ln w.
\label{afpoly}
\end{equation}
Now, it is observed that a solid of hard spheres can tolerate
a polydispersity of approximately 10\%
\cite{pusey87,barrat86,macrae88,bolhuis96} of the
hard sphere diameter.
This makes sense if we note that the lattice
constant of the solid is at most of order 10\% larger than the
diameter of the hard spheres. Hard spheres melt at a fraction
$\simeq 0.74$ of close packing \cite{hoover68} which corresponds
to a lattice constant $\simeq 1.11$ times the hard sphere diameter.
Only if the polydispersity is sufficiently narrow that
few or no spheres are larger than this lattice spacing can the mixture
solidify \cite{pusey87}.
We assume that this is also true for polydisperse sticky spheres;
that they can form a solid lattice if and only if the width of the
polydispersity is less than the difference between the lattice
spacing and the average diameter. Sticky hard spheres solidify into
a solid with a lattice constant $\simeq 1+\delta/2$
and so polydispersity with a width less than $\delta/2$
should allow solidification.
When $w\gg\delta$ the polydisperse spheres are unable to solidify into a single
solid. However, if the polydisperse spheres fractionate, i.e., if they
phase separate into fractions each
with a width $<\delta/2$ then these fractions
can individually solidify. Of course, this phase separation costs
some ideal mixing entropy but at low temperatures this will be outweighed
by the reduction in energy due to the spheres now being in a 12-coordinate
lattice.
The phase separation is to phases with a width $\simeq \delta/2$,
so we have one phase with the spheres in the range of diameters
$(1-w/2)$ to $(1-w/2+\delta/2)$, one phase with the spheres
$(1-w/2+\delta/2)$ to $(1-w/2+\delta)$, etc..
This implies that a fluid with polydispersity of width $w$
solidifies into $2w/\delta$ solid phases.
The free energy of these phases is
equal to that for a monodisperse
solid, Eq. (\ref{as}) minus the mixing entropy for polydisperse spheres
with a distribution of width $\simeq \delta/2$, obtained from
Eq. (\ref{sid}). This is
\begin{equation}
\frac{a_s}{T}= -4\ln\delta -6\frac{\epsilon}{T},
\label{aspoly}
\end{equation}
where we have neglected a term $\ln 2$ as being negligible.

Now we can use the free energies of the fluid and solid phases,
Eqs. (\ref{afpoly}) and (\ref{aspoly}), to determine the
lowest temperature at which the fluid is stable with respect to the solid.
In the limit $\delta\rightarrow0$
\begin{equation}
\frac{a_s-a_f}{T}=\left\{
\begin{array}{ll}
\infty & T>T_{coll}^{poly}\\
-\infty & T<T_{coll}^{poly}\\
\end{array}\right. ,
\label{polylim}
\end{equation}
where $T_{coll}^{poly}$ is
\begin{equation}
\frac{T_{coll}^{poly}}{\epsilon}=\lim\delta\rightarrow 0
\left(\frac{6}{\ln\left(r/\delta^3\right)}\right)~~~~~~\delta^{-1}\gg r\gg 1,
\label{tcollpoly}
\end{equation}
where the lower bound on $r$ ensures that $r$ is large enough that
the number of demixed phases is sufficiently large that
it can be treated as a continuous variable without introducing significant
error, and the upper bound ensures that the polydispersity width
$w\ll 1$.
So, below $T_{coll}^{poly}$ the polydisperse sticky spheres collapse
to form $\simeq 2w/\delta$ solid phases, each with a narrow polydispersity
width of $\simeq \delta/2$. $T_{coll}^{poly}$ is the lowest temperature
at which the fluid phase is stable.
This is the principal result of this work, that
polydispersity delays but does not eliminate the collapse of
sticky spheres, and that the collapse drives fractionation
of the mixture.
We see that as the ratio of the width of polydispersity
to the range of the attraction, $r$, increases the collapse temperature
$T_{coll}^{poly}$ decreases slowly; the variation is only logarithmic.

If we compare Eqs. (\ref{tb20}) and (\ref{tcollpoly}) we
see that as $r$ is at most $\ll\delta^{-1}$, then $T_{coll}^{poly}$ is
always above $T_{B20}$. When the polydispersity of the spheres
is not large in comparison with their diameter then they
always collapse above the temperature at which the
second virial coefficient starts to differ from that of hard
spheres. This contradicts Stell's speculation \cite{stell91}
that polydisperse sticky spheres may show `normal' behaviour,
such as a vapour-liquid transition. We have shown that
this not so when $w\ll 1$.
Actually, we have
implicitly assumed that the thermodynamic limit
is taken before the $\delta\rightarrow0$ limit.
If the order of the two limits is reversed, then as Stell
has shown the dramatic collapse is prevented \cite{stell91}.
Even if we relax the constraint on $w$ and consider
$w={\cal O}(1)$, then so long as $\epsilon$ is the
same for all spheres (which it will not be if the attraction
is due to depletion) then Eq. (\ref{tcollpoly})
still holds and the fluid phase becomes unstable before
the second virial coefficient starts to differ from its hard-sphere
value.
Thus, the phase behaviour first derived by Baxter using the Percus-Yevick
(PY) approximation \cite{baxter68,stell91,barboy74}
is qualitatively incorrect even for polydisperse sticky spheres.

\section{Discussion}

We have found that polydispersity stabilises the fluid phase of sticky
hard spheres. The larger the polydispersity the lower the temperature
at which the fluid phase becomes unstable with respect to
demixing and solidification into a number of coexisting solid phases.
This is a straightforward
consequence of the cost in mixing entropy incurred in phase separating
to form fractions which are sufficiently monodisperse to solidify.
Polydisperse mixtures have been studied before
and the effect of polydispersity on transitions such as solidification
\cite{barrat86,macrae88,bolhuis96}, liquid crystal
transitions \cite{sluckin89} and on transitions in the
fluid state \cite{gualtieri82,sear97} has been studied.
However, there has been no theoretical demonstration of coupled
solidification and phase separation of a
polydisperse mixture. See Ref. \cite{sollich98} for phase
separation of polydisperse mixtures in the fluid phase.

Finally, we comment on the relevance of our findings to demixing in
mixtures of hard spheres. There has been much recent interest in binary
mixtures of hard spheres, see Refs.
\cite{biben96,coussaert97,buhot98,dijkstra98} and references therein.
The question is: Does a binary mixture of small and large hard spheres
ever phase separate to form two coexisting fluid phases?
The answer appears from the latest work \cite{dijkstra98} to be no.
Now, in a mixture of small and large hard spheres the small spheres
can be integrated out, see Refs. \cite{asakura54,vrij76,biben90,dijkstra98}
for details.
Then the binary mixture of hard spheres becomes a single component
system of spheres interacting via a hard core plus
an attraction with a range
of order of the diameters of the small spheres. Thus in the limit
that the ratio $\gamma$
of the diameter of the small spheres to that of the large
spheres is zero the mixture becomes a single component system of sticky
spheres. We know that sticky spheres do not exhibit fluid--fluid
coexistence which implies that a binary mixture of hard spheres does
not exhibit fluid--fluid coexistence in the $\gamma\rightarrow 0$ limit.
This, of course, does not rule out coexistence for
$0<\gamma\ll1$ but Dijkstra {\it et al.} \cite{dijkstra98} find that
fluid--fluid demixing is metastable in this range.
However, the analysis of our section 3 suggests that
even polydisperse sticky spheres do not show fluid--fluid
coexistence.
This implies that a mixture of polydisperse large spheres and
small spheres do not phase separate into two coexisting fluid phases.


It is a pleasure to thank J. Cuesta and D. Frenkel for
useful discussions.

\end{document}